\documentclass[aps,twocolumn,showpacs,preprintnumbers,floatfix,nofootinbib]{revtex4}
\usepackage{graphicx}
\usepackage[subfigure]{graphfig}
\usepackage{epsfig}
\usepackage{epstopdf}
\usepackage{dcolumn}
\usepackage{amsmath}

\def\be{\begin{equation}}
\def\ee{\end{equation}}
\def\bea{\begin{eqnarray}}
\def\eea{\end{eqnarray}}
\newcommand{\nblr}{\overleftrightarrow{D}}

\begin{document}

\title{\vspace{1.0in} {\bf Meson Mass Decomposition from Lattice QCD}} 


\author{Yi-Bo Yang$^{1,2}$, Ying Chen$^{1}$, Terrence  Draper$^{2}$, Ming Gong$^{1,2}$, Keh-Fei Liu$^{2}$,
Zhaofeng Liu$^{1}$, and  Jian-Ping Ma$^{3,4}$
\vspace*{-0.5cm}
\begin{center}
\large{
($\chi$QCD Collaboration)
}
\end{center}
}
\affiliation{
$^{1}$Institute of High Energy Physics, Chinese Academy of Sciences, Beijing 100049, China\\
$^{2}$Department of Physics and Astronomy, University of Kentucky, Lexington, KY 40506\\
$^{3}$State Key Laboratory of Theoretical Physics, Institute of Theoretical Physics, 
Chinese Academy of Sciences, Beijing 100190,  China\\
$^{4}$Center for High Energy Physics, Peking University, Beijing 100871, China 
}

\begin{abstract}
Hadron masses can be decomposed as a sum of quark and glue components which are defined through hadronic matrix elements of QCD operators. The components consist of the quark mass term, the quark energy term, the glue energy term, and the trace anomaly term.  
We calculate these components for mesons with lattice QCD for the first time. 
The calculation is carried out
with overlap fermion on $2+1$ flavor domain-wall fermion gauge configurations. 
We confirm that $\sim 50\%$ of  the light pion mass comes from the quark mass term and $\sim 10\%$ 
comes from the quark energy; whereas, while for the $\rho$ meson, the quark energy 
contributes roughly half of its mass 
but the quark mass term contributes little. 
The combined glue components contribute $\sim 40 - 50\%$ for both mesons. 
It is interesting to observe that the quark mass contribution to the mass of the vector meson is
almost linear in quark mass over a large quark mass region below the charm quark mass. 
For heavy mesons, the quark mass term dominates the masses, while
the contribution from the glue components is about $200$ MeV (a bare value around 2 GeV) for the heavy pseudoscalar and vector mesons. 
The charmonium hyperfine splitting is found to be dominated by the quark energy term which is consistent with the picture of the quark potential model.

\end{abstract}

\pacs{11.15.Ha, 12.38.Gc, 12.39.Mk} \maketitle

\section{Introduction}

Hadrons are confined states of quarks and gluons. QCD is the theory describing the interaction of the quarks and gluons.
Given the fact that masses of hadrons are well measured and successfully calculated with lattice QCD, an interesting, important, and yet unanswered question is how large are the contributions to the masses from
the quark and glue constituents.  The answer will be important for understanding the quark-glue structure of hadrons.
It is clear that the question can only be answered by solving QCD nonperturbatively,
and/or with information from experiment.  The decomposition for the proton has been carried out
with phenomenological inputs~\cite {Ji:1994av}.  For hadrons other than the proton, there is little information from experiments to be used, while some discussion is provided in \cite{Ji:1995pi,Meyer:2007tm}. At the same time, the question can be addressed for all the hadrons
by employing lattice QCD. In this paper, we present such an exploratory study with lattice QCD calculations for the pseudoscalar (PS) and vector (V) mesons.
 
  The energy-momentum tensor from the QCD Lagrangian in Euclidean space~\cite{ccm90} is
\bea \label{eq_EMT}
T_{\mu\nu}=\frac{1}{4}\overline{\psi}\gamma_{(\mu}\nblr_{\nu)}\psi + F_{\mu\alpha} F_{\nu\alpha}
 -  \frac{1}{4}\delta_{\mu\nu}F^2,
\eea
which is symmetric and conserved.  Each term in the tensor depends on the renormalization scale, but the total tensor 
does not. 
The trace term of the tensor is given by
\bea  \label{eq_all_trace}
T_{\mu\mu} = - m \overline{\psi}\psi - \gamma_m m \overline{\psi}\psi + \frac{\beta(g)}{2g} F^2,
\eea
in which the quantum trace anomaly (the term proportional to the anomalous dimension of the mass
operator $\gamma_m$, plus the glue term) has been taken into account. In the above anomaly equation, the first term and the combined second and third terms are scale independent.
The definition of the anomalous dimension of the mass
operator is \cite{DelDebbio:2013sta},
\bea
\gamma_m=-\frac{\mu}{m}\frac{\rm{d} m}{\rm{d}\mu}.
\eea
In lowest order perturbation, the coefficient of the glue anomaly term (the third term in Eq.~(\ref{eq_all_trace})) 
is $\beta(g)=-(11-2n_f/3) g^3/(4\pi)^2$ with $n_f$ being the number of flavors. 

Combining the classical $T_{\mu\nu}$ from Eq.~(\ref{eq_EMT}) and the quantum anomaly in Eq.~(\ref{eq_all_trace}),
one can divide $T_{\mu\nu}$ into a traceless part  $\bar T_{\mu\nu}$ and a trace part $\hat T_{\mu\nu}$, 
i.e. $T_{\mu\nu}= \overline{T}_{\mu\nu} + \hat{T}_{\mu\nu}$ \cite {Ji:1994av}.  
From its matrix element of a single-meson state with momentum $P$,
$\langle P|T_{\mu\nu}|P\rangle=2P_{\mu}P_{\nu}$, and taking $\mu=\nu =4$ in the rest frame, one has
  \bea
 \langle T_{44} \rangle &\equiv& \frac{\langle P|\int d^3 x\, T_{44}(\vec{x})|P\rangle}
 {\langle P|P\rangle} = - M, 
\nonumber\\
 \langle \overline{T}_{44} \rangle &=& - 3/4 M, \,\,\,\,\,
 \langle \hat{T}_{44} \rangle = - 1/4 M,
 \eea
  with
 \bea
\overline{T}_{44} &=&\frac{1}{4}\overline{\psi}\gamma_{(4}\nblr_{4)}\psi-\frac{1}{16}\overline{\psi}\gamma_{(\mu}\nblr_{\mu)}\psi\nonumber\\
&&+F_{4\alpha} F_{4\alpha}
 -  \frac{1}{4}F^2\nonumber\\
 &=&\sum_{u,d,s}(\overline{\psi}\gamma_{4}\overrightarrow{D}_{4}\psi+ \frac{1}{4}m\overline{\psi}\psi) + \frac{1}{2}(E^2-B^2),\\
 \hat{T}_{44} &=& \frac{1}{4}T_{\mu\mu}\nonumber\\
 &=&\frac{1}{4}\lbrack- (1+\gamma_m) \sum_{u,d,s...}m \overline{\psi}\psi + \frac{\beta(g)}{g} (E^2+B^2)\rbrack,\nonumber\\
 \eea
for the zero momentum case.
 The Hamiltonian of QCD can be decomposed as~\cite{Ji:1994av}
 \bea
 H_{QCD} &\equiv& - \int d^3 x\, T_{44} (\vec{x}) = H_q  + H_g + H_g^a + H^{\gamma}_m,\\
 H_q &=& - \sum_{u,d,s...}\int d^3 x~  \overline \psi(D_4\gamma_4)\psi, 
 \nonumber\\
 H_g &=& \int d^3 x~ {\frac{1}{2}}(B^2- E^2), 
 \nonumber\\
 H_g^a &=&\int d^3x~ \frac{-\beta(g)}{4g}( E^2+ B^2),\nonumber\\
H^{\gamma}_m &=&\sum_{u,d,s\cdots}\int d^3x\, \frac{1}{4}\gamma_m m\, \overline \psi  \psi
 \eea
with $H_q$, $H_g$, $H^a_g$, and $H^{\gamma}_m$ denoting the total contributions from the quarks, the glue field energy, the QCD glue trace anomaly, and the quark mass anomaly, respectively. 
Note that  the sum of the first two and the sum of the last two terms are separately scale and renormalization scheme independent, while each term separately is not.
Using the equation of motion (EOM), $H_q$ can be further divided into quark energy and mass terms
\bea   \label{EOM}
 H_q = H_E + H_m,
 \eea
 with 
 \bea
 H_E &=& \sum_{u,d,s...}\int d^3x~\overline \psi(\vec{D}\cdot \vec{\gamma})\psi, 
 \nonumber\\
 H_m &=& \sum_{u,d,s\cdots}\int d^3x\, m\, \overline \psi  \psi.  
 \eea
N.B.: the quark energy $H_E$ includes both kinetic and potential energies due to the covariant derivative.
Given the above division, a hadron mass can be decomposed into the following matrix elements,
\bea
M &=& - \langle T_{44} \rangle= \langle H_q \rangle + \langle H_g\rangle + \langle H_a\rangle + \langle H^{\gamma}_m\rangle
\nonumber\\
&=&\langle H_E\rangle + \langle H_m\rangle+ \langle H_g\rangle +  \langle H_a \rangle, 
\label{eq:T44}\\
\frac{1}{4}M &=& -\langle \hat{T}_{44} \rangle= \frac{1}{4}\langle H_m\rangle + \langle H_a\rangle,
\label{eq:trace}
\eea
with  all the $\langle H \rangle$ defined by $\langle P|H |P\rangle/\langle P|P\rangle$ and 
\bea
\langle H_a\rangle=\langle H^{\gamma}_m\rangle+\langle H^a_g \rangle
\eea
as the total trace anomaly.
Each matrix element can be calculated with lattice QCD. 
Since hadron masses can be 
obtained from the two-point correlators on the lattice, we shall calculate 
$\langle H_q\rangle$ (or $\langle H_E\rangle$) and $\langle H_m \rangle$ through the three-point correlators and extract
 $\langle H_a\rangle$ and $\langle H_g \rangle$ from Eqs.~(\ref{eq:T44}) and (\ref{eq:trace}) in
this work. We will directly calculate these glue matrix elements in the future. 

The structure of the rest of the paper is organized as follows. The numerical details of the simulation, including the fermion action and configurations used, and the systematic uncertainties, will be discussed in Sec.~\ref{sec:numerical}. In Sec.~\ref{sec:result}, the results such as the condensates in the mesons, the decomposition of the PS/V mesons and their difference (the splitting) are provided. A short summary and outlook are presented in Sec.~\ref{sec:summary}.

\section{Numerical details}\label{sec:numerical}

\par 

In this work, we use the valence overlap fermion on $2 +1$ flavor domain-wall fermion (DWF) configurations \cite{Aoki:2010dy} to carry out the calculation~\cite{Li:2010pw}. The effective quark propagator of the massive
overlap fermion is the inverse of the operator $(D_c + m)$~\cite{Chiu:1998eu,Liu:2002qu}, where  $D_c$ is chiral, i.e. $\{D_c, \gamma_5\} = 0$ \cite{Chiu:1998gp}, and is expressed in terms of the overlap operator $D_{ov}$ as
\bea
D_c=\frac{\rho D_{ov}}{1-D_{ov}/2} \textrm{ with }D_{ov}=1+\gamma_5\epsilon(\gamma_5D_{\rm w}(\rho)),
\eea
where $\epsilon$ is the matrix sign function and $D_{\rm w}$ is the Wilson Dirac operator with a negative mass
characterized by the parameter $\rho=4-1/2\kappa$ for $\kappa_c < \kappa < 0.25$. We set $\kappa$=0.2  which corresponds to $\rho=1.5$. 

\begin{figure}[b]
  \includegraphics[scale=0.6]{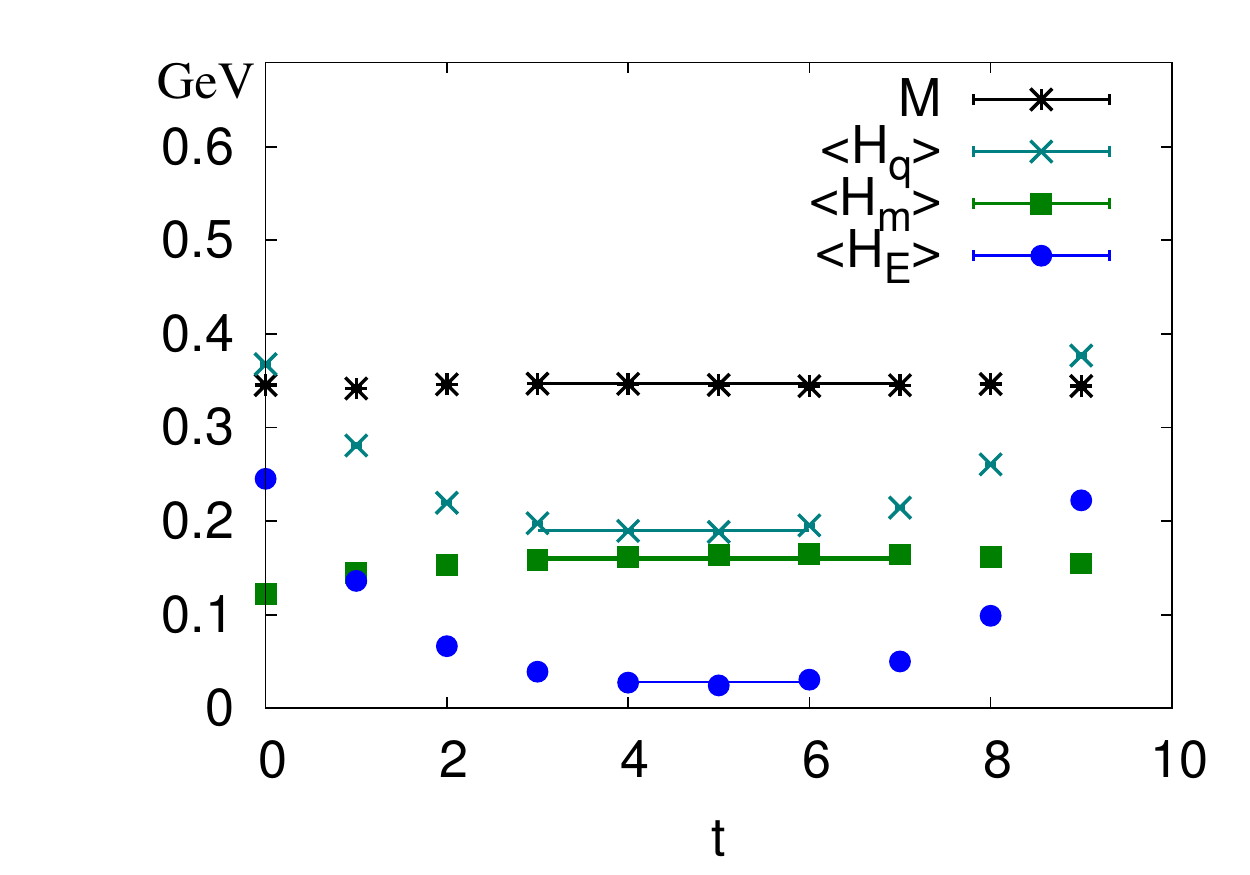} 
 \includegraphics[scale=0.6]{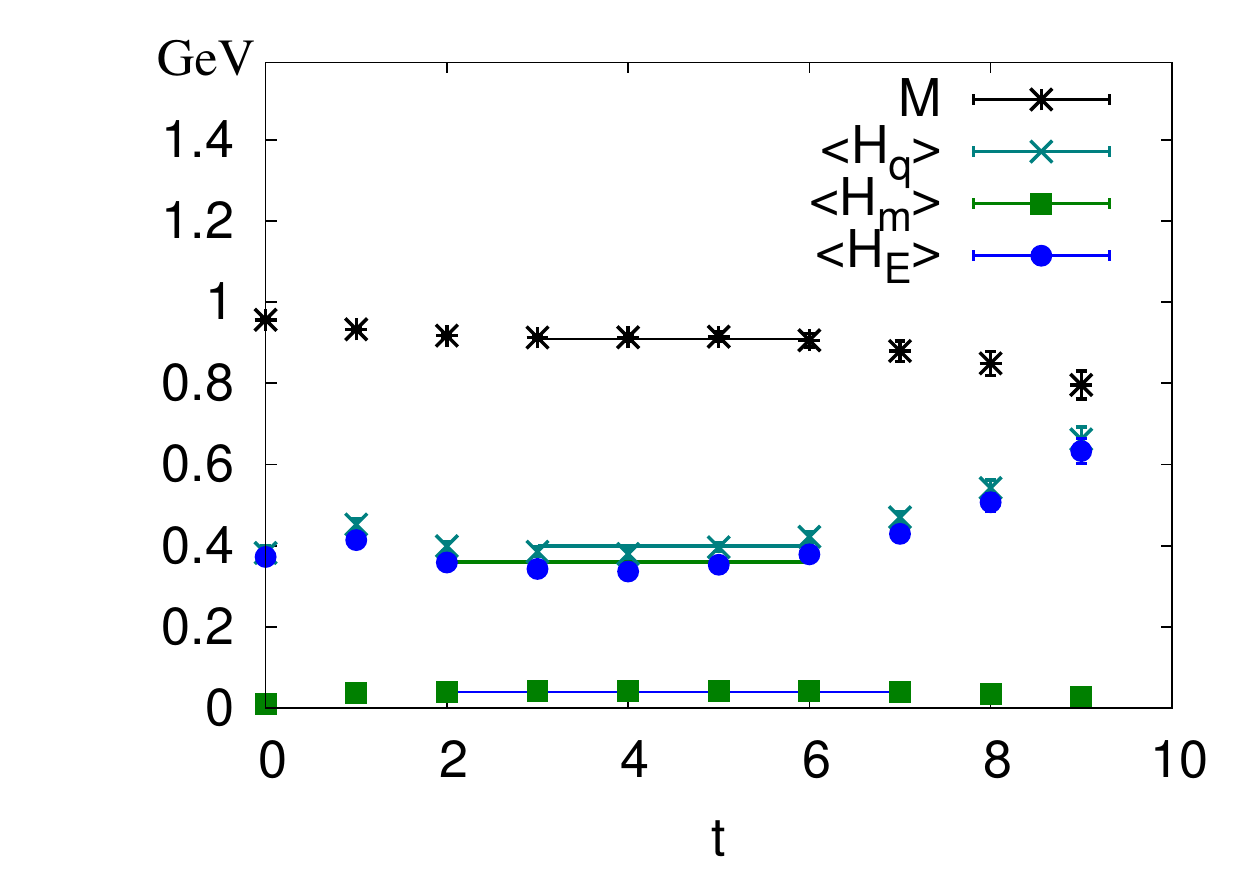} 
 \caption{Plateaus of quark components of (a) PS mesons and (b) V mesons with light quark pairs which correspond to $m_{\pi}\sim$ 330 MeV. Half of each of the light PS/V meson masses comes from the glue, while the other half is dominated by the quark mass/energy in the PS/V case respectively.}\label{fig:mass}
\end{figure}

The lattice we use has a size $24^3\times 64$ with lattice spacing $a^{-1} = 1.77(5)$ GeV set by Ref.~\cite{Yang:2013gf}. The light
sea $u/d$ quark mass $m_{l}a = 0.005$ corresponds to $m_{\pi} \sim 330$ MeV. We have calculated
the PS and V meson masses and the corresponding $\langle H_m\rangle, \langle H_q\rangle$,
and $\langle H_E\rangle$ at 12 valence quark mass parameters which correspond
to the renormalized masses $m_q^R\equiv m_q^{\overline{\rm MS}}(2 \rm GeV)$ ranging from 0.016 to 1.1 GeV after
the non-perturbative renormalization procedure in Ref.~\cite{liu:2013renorm}. The smallest one is slightly smaller than the sea quark mass and corresponds to a pion
mass at 281 MeV, and the largest quark mass is close to that of the charm.
In order to enhance the signal-to-noise ratio in the calculation of three-point functions, we set two
smeared noise grid sources at $t_i=0/32$ \cite{Gong:2013vja} and four noise-grid point sources at positions $t_f$ which are 10 time-slices away from the sources on 101 configurations. To obtain a better signal in the light quark region ($<$0.1 GeV), the low mode substitution technique \cite{Li:2010pw} is applied to the contraction in those cases.

It would be ideal to use the conserved lattice stress tensor and there are attempts to construct the conserved stress tensor on the lattice perturbatively and non-perturbatively \cite{ccm90} and recently by Suzuki \cite{Suzuki:2013gza,Makino:2014taa} with Wilson flow at finite lattice spacing. However, these approaches inevitably involve complicated sets of operators, which are difficult to compute in the lattice calculation. Our approach is to use the quark stress operators with lattice derivative (the point-split operators with gauge links \cite{ccm90}) for $\overline{\psi}\gamma_{\mu}\nblr_{\nu}\psi$,
\bea
&\!\!\!\frac{1}{2}&\!\!\!\big(\overline{\psi}(x)\gamma_{\mu}(U_{\nu}(x)\psi(x+\hat{\nu})-U_{\nu}^{\dagger}(x-\hat{\nu})\psi(x-\hat{\nu}))\nonumber\\
&\!\!\!+&\!\!\!(\overline{\psi}(x+\hat{\nu})U_{\nu}(x+\hat{\nu})-\overline{\psi}(x+\hat{\nu})U_{\nu}^{\dagger}(x))\gamma_{\mu}\psi(x)\big)\nonumber\\
&\!\!\!=&\!\!\! a \gamma_{\mu}(x)\nblr_{\nu}\psi(x) +O(a^3),
\eea
 to carry out lattice calculation at a finite cutoff and then to extrapolate to the continuum limit as a next step. 

The matrix elements for the operators $\overline{\psi}\gamma_4\nblr_4\psi, \overline{\psi}\gamma_i\nblr_i\psi$ and $m \overline{\psi}\psi$ are extracted from the plateaus of the ratios of three-to-two point functions to obtain $\langle H_m\rangle, \langle H_q\rangle$, and $\langle H_E\rangle$ in the connected insertions for different quark masses. In the present work, we only consider the equal-mass case for the quark-antiquark pairs in  the mesons. 

We show in Fig.~\ref{fig:mass} the ratio of three- to two-point functions for (a) 
PS mesons and (b) V mesons with light quark pairs, which corresponds to $m_{\pi}\sim$ 330 MeV. We see that 
the plateaus for  $\langle H_m\rangle, \langle H_q\rangle$, $\langle H_E\rangle$ from the ratio of three-to-two point functions are clearly visible. 
At the same time, the plateaus of the total mass $M$ from the effective mass of the two point function with 
noise-smeared grid source are also long enough to obtain precise results.
We also applied a curve fit including the contribution 
of excited states to extract the matrix elements and found that the results are consistent with the ones from the constant fit.

As observed in Fig.~\ref{fig:mass}, the quark mass term $\langle H_m\rangle$
contributes about half of the light PS mass, while the quark energy term 
$\langle H_E\rangle$ is very small.  This implies that the other half of the light PS mass comes 
mainly from the glue. For the light V mass, the combined glue components also contributes roughly one half, 
while $\langle H_E\rangle$ is dominant in the other half, and $\langle H_m\rangle$ is small.

\subsection{Equation of motion and quark energy}

 Before presenting our results, we discuss the theoretical underpinning of the equation of motion (EOM)
in the context of lattice calculation of three-point functions. In the three-point function with the operator 
$D_c + m$ inserted
at a time different from the meson source and sink, part of the correlator will involve the
product of the operator and a quark propagator which has the relation
\be
\sum_z (D_c +m)_{(x,z)}.\frac{1}{D_c+m}_{(z,y)}=\delta_{x,y},
\ee
where $x,y,z$ denote all the space-time, color and Dirac indices. Since the inserted operator $D_c + m$ is at a different
time from that of the source time, $x \neq y$. As a result, the matrix element of $D_c + m$ is zero.
For the disconnected insertion (DI), the delta function leads to a constant for the quark loop. Since the uncorrelated part after gauge averaging is to be subtracted, this also gives a null result for $D_c + m$ in the DI. 
Therefore, the matrix element with the insertion of the $D_c + m$ operator is zero which is just the  
equation of motion on the lattice for fermions with the quark mass as an additive constant in the fermion
propagator. This does not hold straightforwardly for the Wilson fermion where there is an additive mass renormalization and mixing with lower dimensional operators which need to be taken into account. 

\par 
Since $D_{ov}$ has eigenvalues on a unit circle centered at 1 on the
real axis, the eigenvalues of $D_c$ are purely imaginary except the zero modes~\cite{Liu:2002qu}. This is the same as in the continuum. Thus, $\overline{\psi}D_c\psi$ approaches $\overline{\psi}\gamma_{\mu}D_{\mu}\psi$ with an $\mathcal{O}(a^2)$ error and we thus have
$\langle H_q \rangle = \langle H_E \rangle + \langle H_m\rangle + O(a^2)$ as in the 
continuum in Eq.~(\ref{EOM}) modulo an $\mathcal{O}(a^2)$ error.

\begin{figure}[htb]
\centering
\includegraphics[scale=0.6]{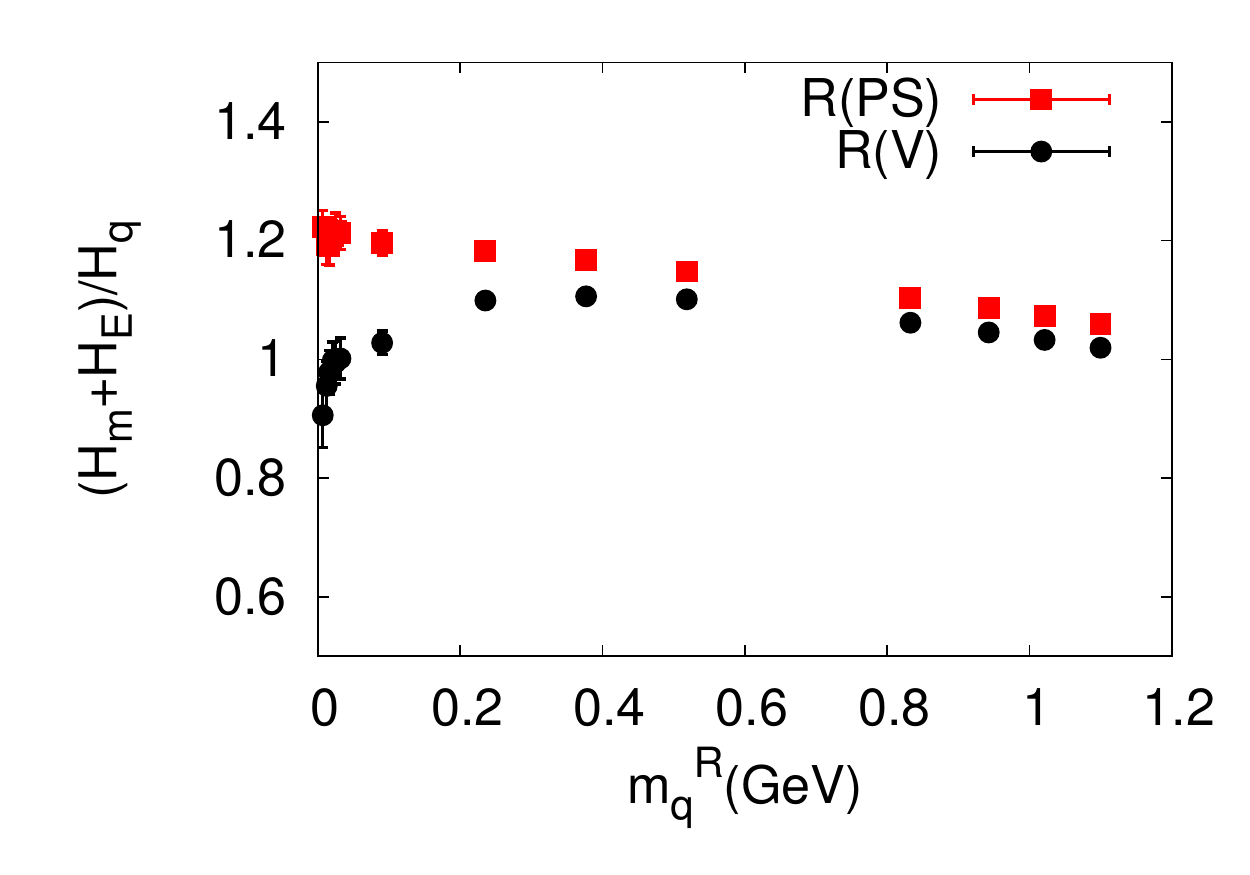}
\vspace*{-0.5cm}
 \caption{The ratio $R$ as defined in the text  as a function of the renormalized quark mass
 $m_q^R$ in the $\overline{MS}$ scheme at 2 GeV, to estimate the $O(a^2)$ error due to the breaking of EOM. The systematic uncertainty vanishs when $R=1$. The values of $R$ are close to unity except for the very light 
 quark mass region in the PS meson case.}
\label{fig:renorm}
\end{figure}

One can estimate this $\mathcal{O}(a^2)$ error by considering the ratio
of  $\langle D_c\rangle$ to $\langle H_E -  H_q\rangle$. Both matrix elements approach the same  matrix element
of  $\overline{\psi}\gamma_{\mu}\nblr_{\mu}\psi$ in the continuum. This is equivalent to
considering the ratio 
\bea
R = \frac{\langle H_m\rangle + \langle H_E\rangle }{\langle H_q\rangle},
\eea
which should be equal to $1 + \mathcal{O}(a^2)$.
We plot $R$ as a function of the quark mass in Fig.~\ref{fig:renorm} for the PS and V mesons. 
Except for the region of  very light quark masses, they are roughly the same for PS and V mesons. 
In the charm quark mass region, the ratio is close to unity, while for light quarks, the ratio for the
PS can be as large as $\sim 1.2$ and it is close to unity for the V mesons. We shall take 20\% as 
an conservative estimate for the systematic errors of $\langle H_E\rangle$ and $\langle H_q\rangle$ for
the light quark case due to the finite lattice spacing.

\subsection{Disconnected insertion}

The DI contribution to the quark mass and quark energy terms needs to
be estimated stochastically which is usually quite a bit noisier than that of the connected insertion (CI).  It is found recently that the signal of the DI contribution for the quark loops with scalar density can be highly improved by the low-mode averaging (LMA) in the loop and low-mode substitution for the nucleon propagator~\cite{Gong:2013vja}. We use this approach and calculate the DI of the quark mass term for the PS and V mesons to gauge the DI 
contributions.

\begin{figure}[b]
  \includegraphics[scale=0.6]{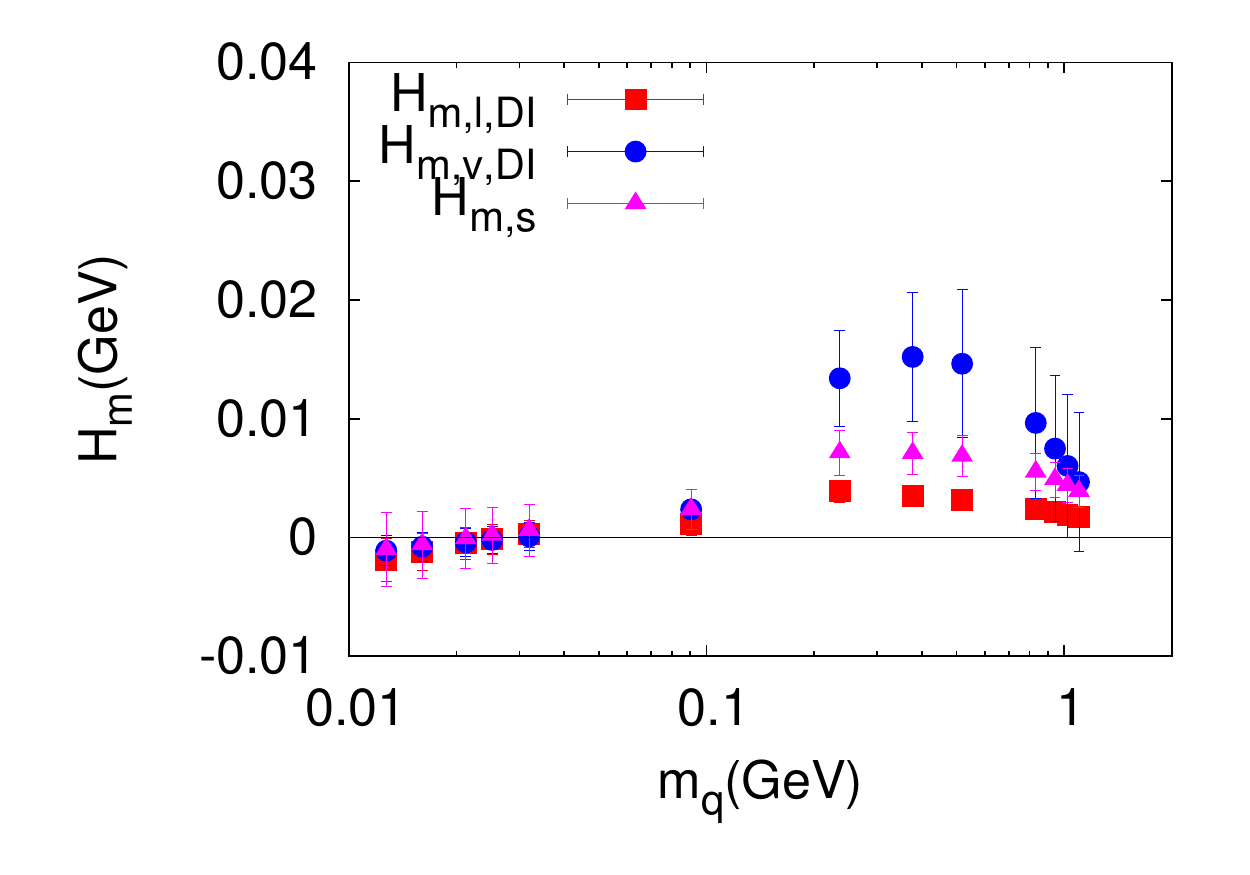} 
 \includegraphics[scale=0.6]{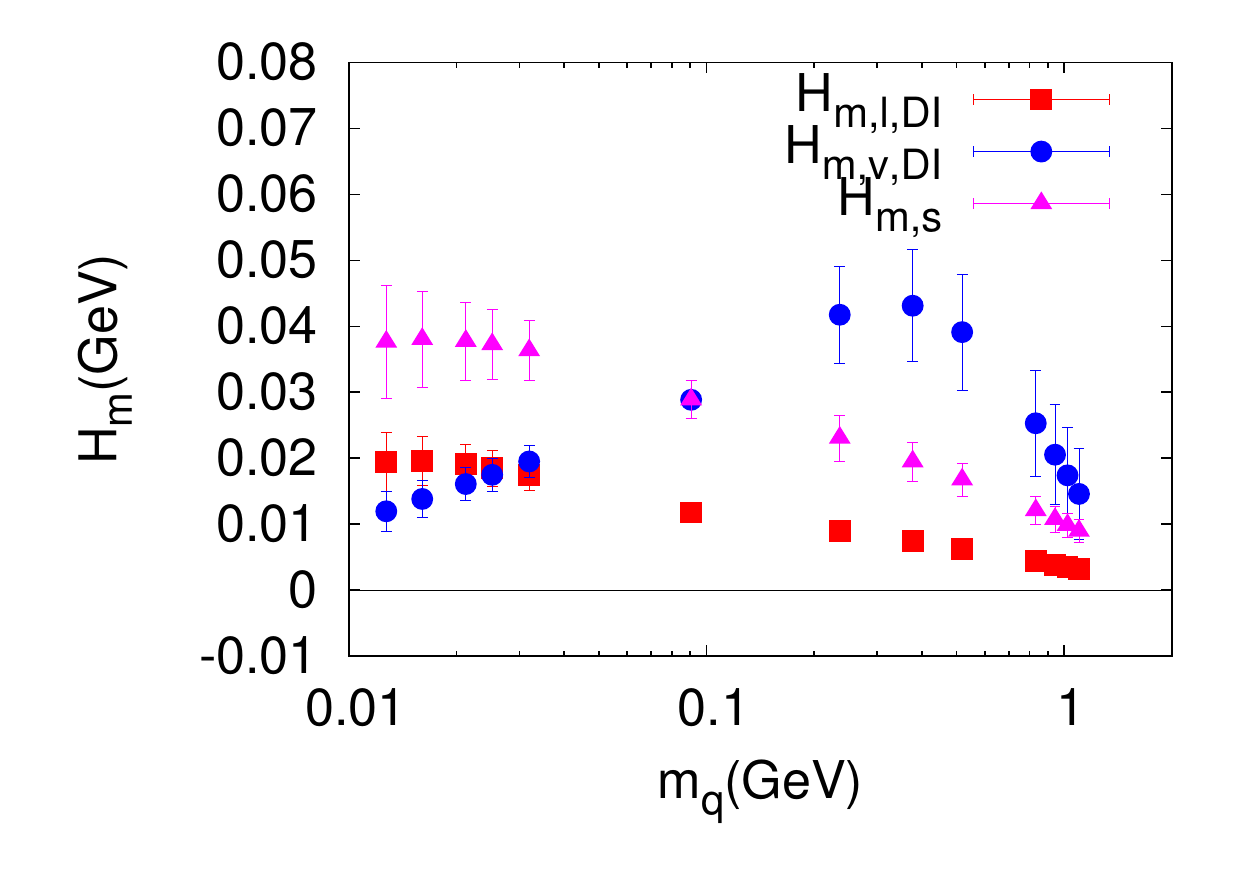} 
 \caption{The contributions of the quark mass term in the PS/V meson from kinds of DI diagrams. The red-square, blue-circle, and pink triangle points are the contributions from the quark loop with mass equal to light sea quark mass, valence quark mass, and strange sea quark mass correspondingly. It is easy to confirm that the DI contribution to the quark mass term in the PS meson case is just a few MeV for all cases, while that in the V meson case is larger but no more than $\sim 40$ MeV for all the cases.}\label{fig:H_m_di}
\end{figure}

   We show the $H_m$ results for the DI in Fig.~\ref{fig:H_m_di} for the cases where the quark loop mass
equals to that of the light sea (i.e.\ corresponding to $m_{\pi} = 330$ MeV), the valence quark mass in the
meson, and the strange mass as a function of the valence quark mass of the meson. We see in the upper panel of Fig.~\ref{fig:H_m_di} that in all cases, the DI contribution of $H_m$ is of the order of a few MeV for all the PS mesons. For the case of the V mesons in the lower panel of  Fig.~\ref{fig:H_m_di}, they are also
small with the largest contribution being $\sim 40$ MeV due to the strange quark contribution in the light
quark case.  It is also about 40 MeV for the case where the the loop and valence has the same quark mass for
$m_q > 0.2$ GeV where the vector meson mass is greater than 1.5 GeV.  Based on these
estimates, we shall ignore the small and noisy DI contribution of $H_m$ in this study. 

   In view of the fact the strange momentum fraction $\langle x\rangle_s$ is only $2-4$\% of the total nucleon
momentum experimentally~\cite{Olness:2003wz} and the ratio of   $\langle x\rangle_s$  to 
 $\langle x\rangle_{u/d}$ in the DI is 0.78(3) from the latest lattice calculation with overlap fermion on
 DWF configurations on the $24^3 \times 64$ lattice~\cite{Sun}, we expect the DI contribution of
 the quark energy term is an order of  magnetic smaller than that of the CI. Thus, we shall neglect them
 also in the present study.

\subsection{The scale and renormalization scheme dependence of the decomposition}

In the decomposition equation 
\bea
M &=&\langle H_a\rangle+\langle H_m\rangle+ \langle H_g\rangle +  \langle H_E \rangle, 
\label{eq:T44_new}
\eea
each of the first two terms and the sum of the last two terms are separately scale and renormalization scheme independent, while each of the last two terms is not. 

Another way to decompose the first two terms is to absorb the quark mass anomaly term $\langle H^{\gamma}_m\rangle$ into the quark mass term $\langle \bar{H}_m\rangle=(1+\frac{1}{4}\gamma_m)\langle H_m\rangle$, leaving the QCD glue trace anomaly $\langle H^a_g\rangle$ alone. Using Eqs.~(46-49) of Ref.~\cite{Chetyrkin:1999pq}, we would get the 4-loop result of $\gamma_m$ in the $\overline{MS}$ scheme at 2GeV to be,
\bea
\gamma_m^{\overline{MS}}(2\,\textrm{GeV})=0.26(1).
\eea
with the uncertainty estimated by its 4th order contribution. The higher order corrections of $\beta(g)$ could also be found in Ref.~\cite{Chetyrkin:1999pq}, but this would be beyond the demand of this work since we don't calculate the glue contribution directly. Such a decomposition is not favored, however, as both the total quark mass term and the QCD glue anomaly term depend on both renormalization scheme and scale. It is worthwhile to note that, in the light quark case, the anomaly quark mass term is suppressed by both the quark mass and the factor $\gamma_m/4$ so that the two kinds of decompositions are not significantly different numerically.

For the quark/glue energy, the present calculation of the quark energy is its lattice value at the scale around the inverse of the lattice spacing $a^{-1}=1.77$GeV. Since the combined energy is scale and renormalization scheme independent, the glue energy is also at the same scale. 
In principle, to obtain the result at 2GeV in the 
$\overline{MS}$ scheme, we could calculate the renormalization of the quark energy in RI/MOM scheme with simulation and then convert it into $\overline{MS}$ scheme perturbatively, or use lattice perturbation theory to calculate the renormalization of the quark energy in $\overline{MS}$ with finite lattice spacing. Also, the mixing effect between the quark/glue components should be taken into account \cite{Meyer:2007tm,Deka:2013zha}.

\section{Results}\label{sec:result}

\subsection{The scalar matrix element in the PS/V meson}  \label{condensate}

\begin{figure}[htb]  
\centering
\includegraphics[scale=0.6]{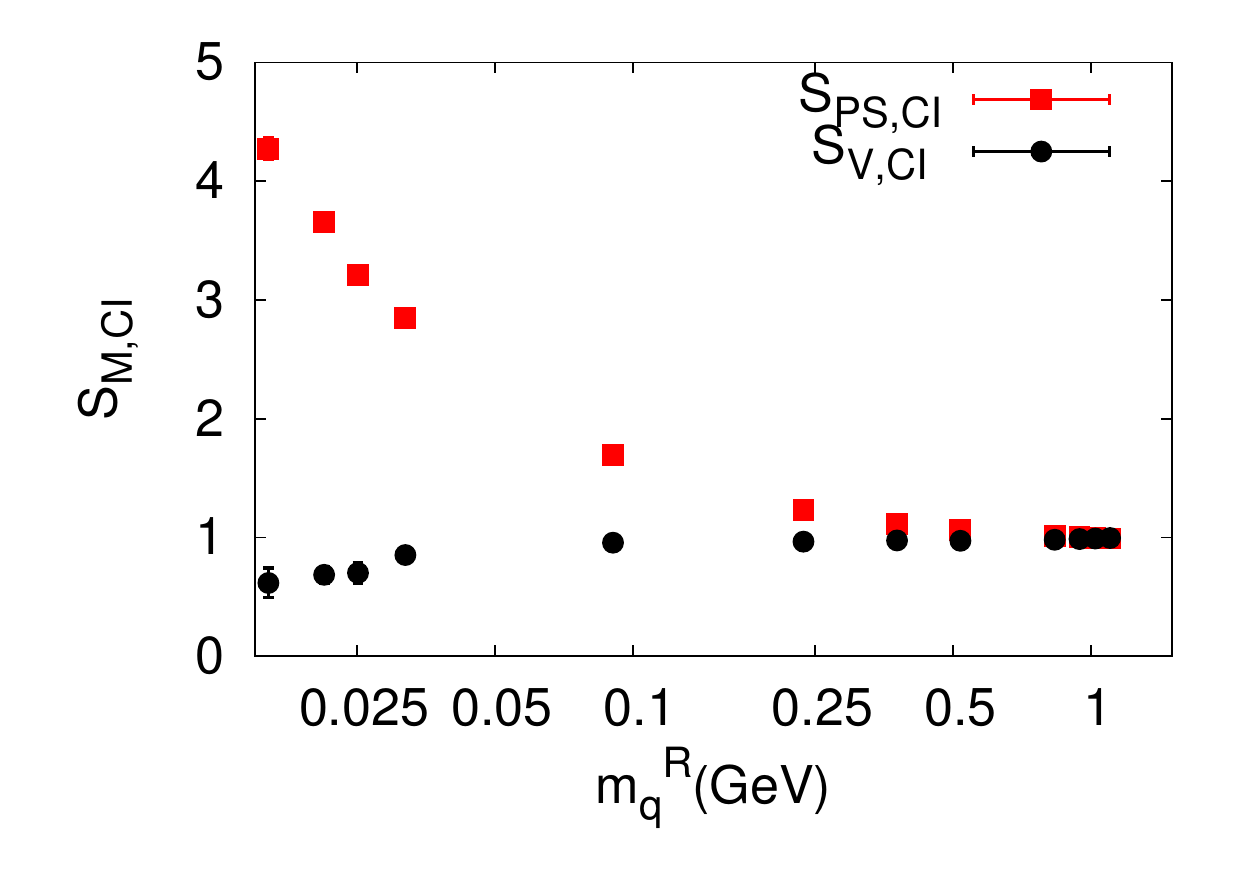}
\vspace*{-0.5cm}
 \caption{Matrix elements $S_{\rm M, CI}$ for the PS and V mesons. 
 $S_{\rm PS, CI}$ (red square) increases with decreasing $m_q^{R}$ while $S_{\rm V, CI}$ (blue circle)
 remains constant and is close to unity throughout the quark mass range below the charm quark mass.}
\label{fig:scalar}
\end{figure}

The CI parts of the matrix element \mbox{$S_{\rm M}\equiv \langle M|\int d^3x\, \overline{\psi}\psi|M\rangle/\langle M|M\rangle$} for PS and V are
plotted in Fig.~\ref{fig:scalar}. Even though they are almost identical in the heavy quark region as expected, $S_{\rm PS, CI}$ increases with decreasing $m_q^{R}$, while $S_{\rm V, CI}$ remains largely constant and is close to unity throughout the quark mass range below the charm quark mass. 
From the Feynman-Hellman theorem, 
\bea
S_{\rm M, CI}= \frac{\partial M}{\partial m_{v}} \label{eq:Feynman-Hellman_CI},
\eea
one can easily deduce the $1/\sqrt{m_q^R}$ behavior of PS mesons with the Gell-Mann-Oakes-Renner relation 
\mbox{$m_{PS}^2 = - 2 m_q \langle \bar{q}q \rangle/f_{\pi}^2$}. We will see later that Eq.~(\ref{eq:Feynman-Hellman_CI}) is also useful to understand the quark mass dependence of the V meson.

\subsection{Pseudoscalar meson}

\begin{figure}[htb]
  \includegraphics[scale=0.6]{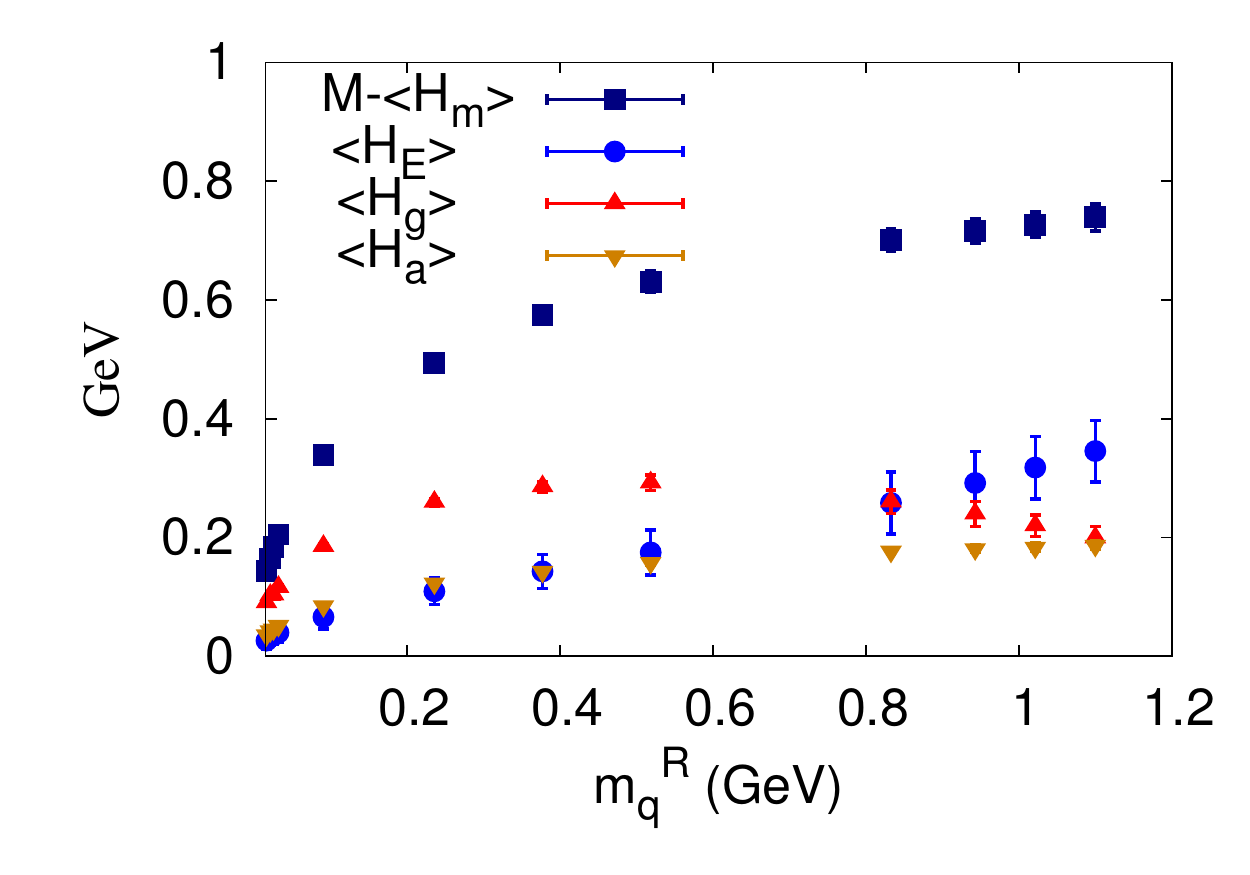} 
\caption{Different component contributions to the PS mass as functions of the renormalized valence quark mass. The contributions from the glue energy and trace anomaly are stable in the region from 0.2 to 0.8 GeV, while the former one decreases for heavier quark masses.}\label{fig:glue_ps1}

\end{figure} 
\begin{figure}[htb]  
 \includegraphics[scale=0.6]{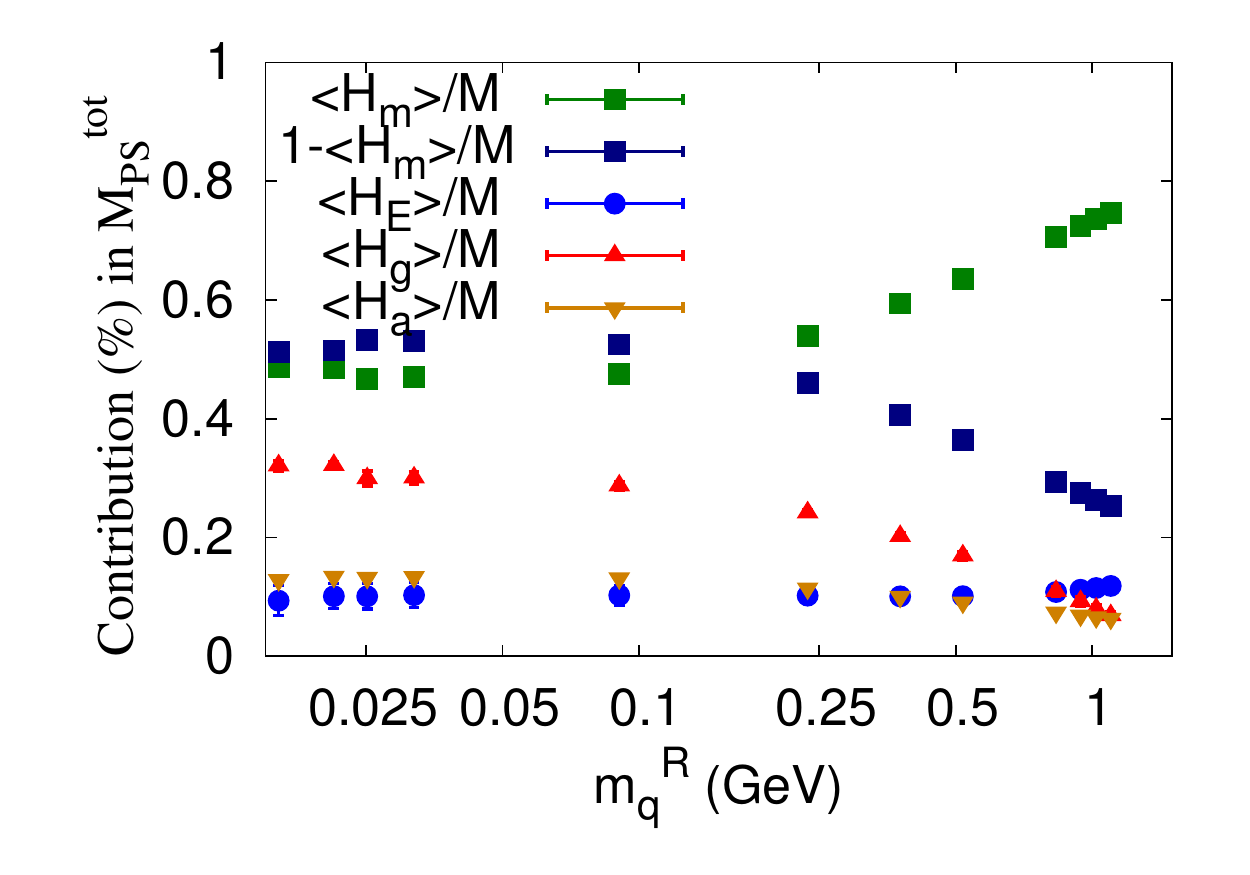} 
\caption{Ratios of the different component contributions to the PS mass as functions of the renormalized valence quark mass. All these ratios are positive which suggests that all the components approach zero at the chiral limit.}\label{fig:glue_ps2}
\end{figure}

    Our lattice results of the difference between $M$ and the quark mass term $\langle H_m\rangle$, i.e. $M - \langle  H_m\rangle$, the quark
kinetic and potential energy term $\langle H_E\rangle$, the glue field energy $\langle H_g\rangle$, and
the anomaly $\langle H_a\rangle$ for the PS meson as a function of the renormalized
valence quark mass are presented in Fig.~\ref{fig:glue_ps1}. It is interesting to observe that all these contributions are positive which suggests that they all approach zero at the chiral limit when the pion mass approaches zero. We also note that the contributions from the glue energy and trace anomaly are stable in the region of $m_q^R$  from 0.2 to 0.8 GeV and then the former one decreases for the heavier quark mass
case.

We also plot the ratios of the quark and glue components with respect to $M$ in Fig.~\ref{fig:glue_ps2}.
For the light PS mesons, the quark mass term is about 50\% of the total mass.
This can be derived from the Gell-Mann-Oakes-Renner relation $m_{\pi} \propto \sqrt{m^q}$ and
the Feynman-Hellman relation Eq.~(\ref{eq:Feynman-Hellman_CI}).
 This implies from Eq.~(\ref{eq:trace}) that the anomaly term $\langle H_a\rangle$ 
contributes $\sim 12\%$ of the mass. 
The remaining contributions from $\langle H_g\rangle$ and $\langle H_E\rangle$ are $\sim 30\%$ and
$\sim 8\%$ respectively.  They are consistent with an estimate based on the quark momentum fraction
in the pion and the Gell-Mann-Oakes-Renner relation~\cite{Ji:1995pi}. Since
our present results are from the partially quenched calculation, it will be interesting
to check them again in the future for configurations with physical light sea quark masses.

\subsection{Vector meson}

\begin{figure}[htb]
 \includegraphics[scale=0.6]{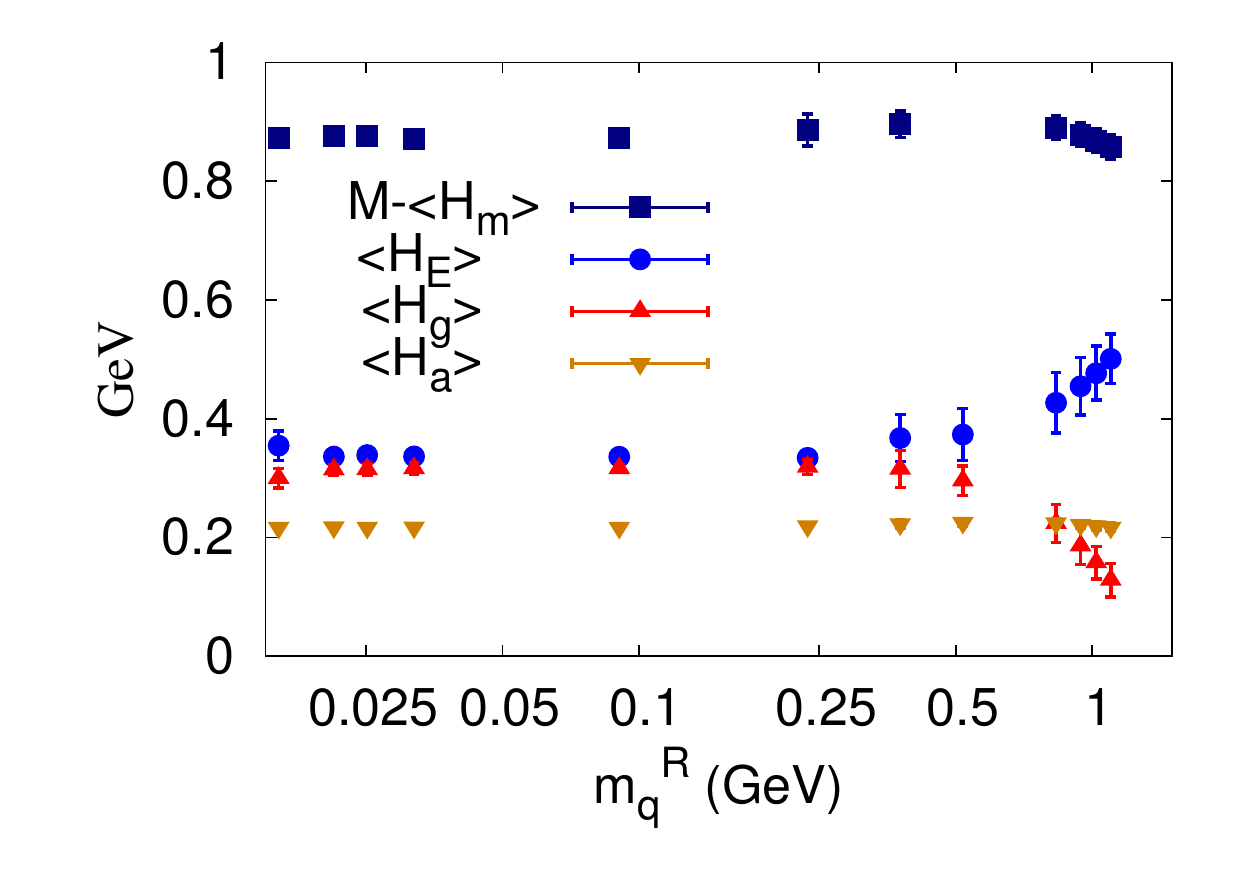}
\caption{Same as in Fig.~\ref{fig:glue_ps1} for the vector meson. The contributions from all the components except the quark mass term (which will be discussed in Fig.~\ref{fig:glue_v1.5}) are stable when the valence quark mass is smaller then $\sim 500$ MeV.}
\label{fig:glue_v1}
\end{figure}

\begin{figure}[htb]
 \includegraphics[scale=0.6]{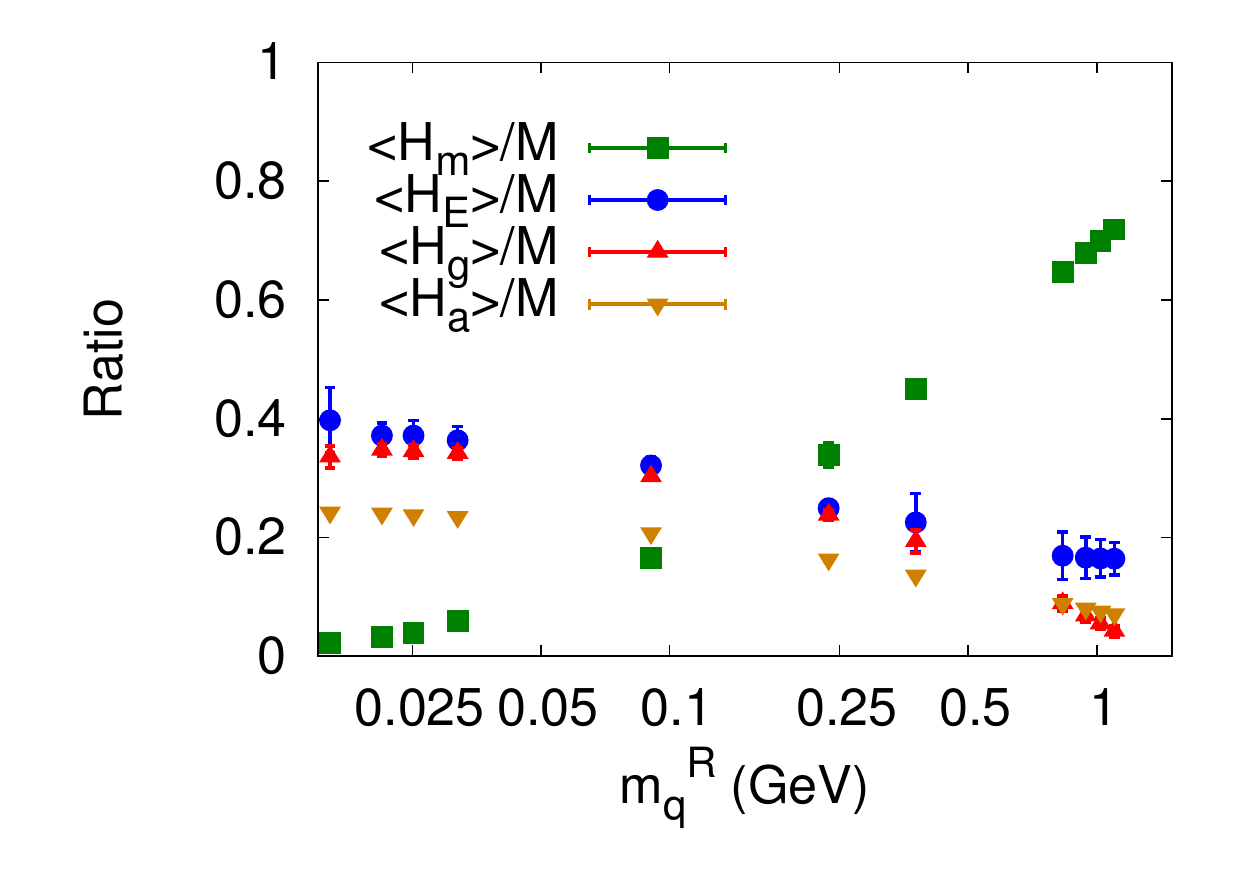}
\caption{Same as in Fig.~\ref{fig:glue_ps2} for the vector meson. The quark mass contribution (blue square) increases from almost zero in the light quark region to over 80\% in the charm quark region.}
\label{fig:glue_v1.5}
\end{figure}

The same components in the V mesons and their ratios to the total mass are plotted in Figs.~\ref{fig:glue_v1} 
and \ref{fig:glue_v1.5}. Close to the chiral limit, $\langle H_E\rangle$ constitutes $\sim 40\%$ of the $\rho$ meson mass, while the sum of the glue energy and anomaly terms contribute about 60\% and $\langle H_m\rangle$ vanishes like $O(m_q^R)$. 

\begin{figure}[htb]
 \includegraphics[scale=0.6]{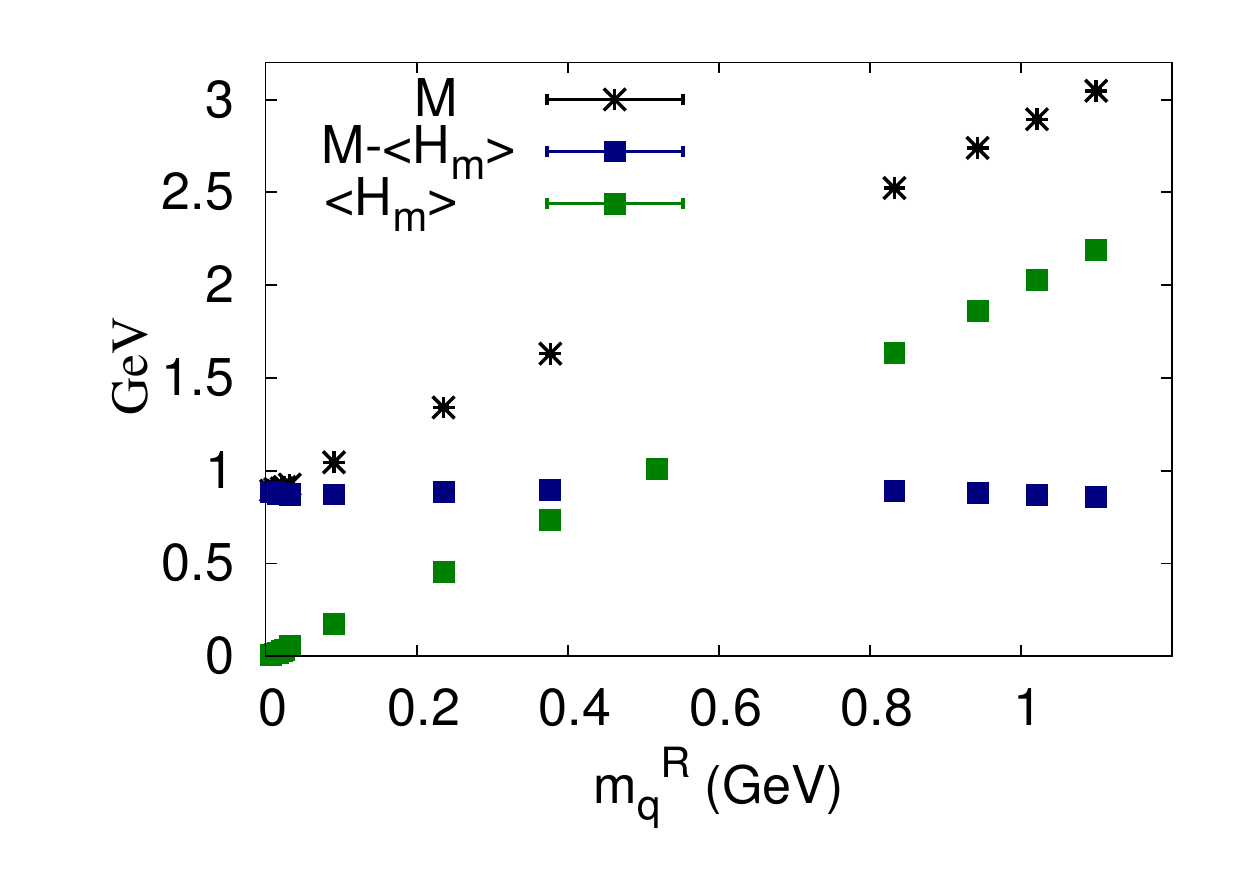}
\caption{The quark mass dependence of the V meson mass is linear in the current quark mass and comes mostly from $\langle H_m\rangle$; see their difference marked by the dark blue square.}
\label{fig:glue_v2}
\end{figure}

For the heavier V mesons, the behavior 
\bea
M_V(m_q^{R})\sim2 m_q^{R} C_0+\textrm{const.} 
\eea
with $C_0$ a constant is observed in Fig.~\ref{fig:glue_v2}. 

We note that the components $\langle H_E\rangle$, $\langle H_g \rangle$ and
$\langle H_a\rangle$ are insensitive to the current quark mass through the quark
mass region less than $\sim 500$ MeV. In this region, the glue energy and trace anomaly contribution to the V meson mass, i.e.  
$\langle H_a \rangle +\langle H_g\rangle$, is about 500 MeV and the quark energy $\langle H_E \rangle$ 
contributes about 350 MeV. It is tantalizing to consider the possibility that the
constant glue contribution and quark energy could be the origin of the constituent quark mass in the quark model
picture.

\subsection{Hyperfine splitting}

\begin{figure}[]
  \includegraphics[scale=0.6]{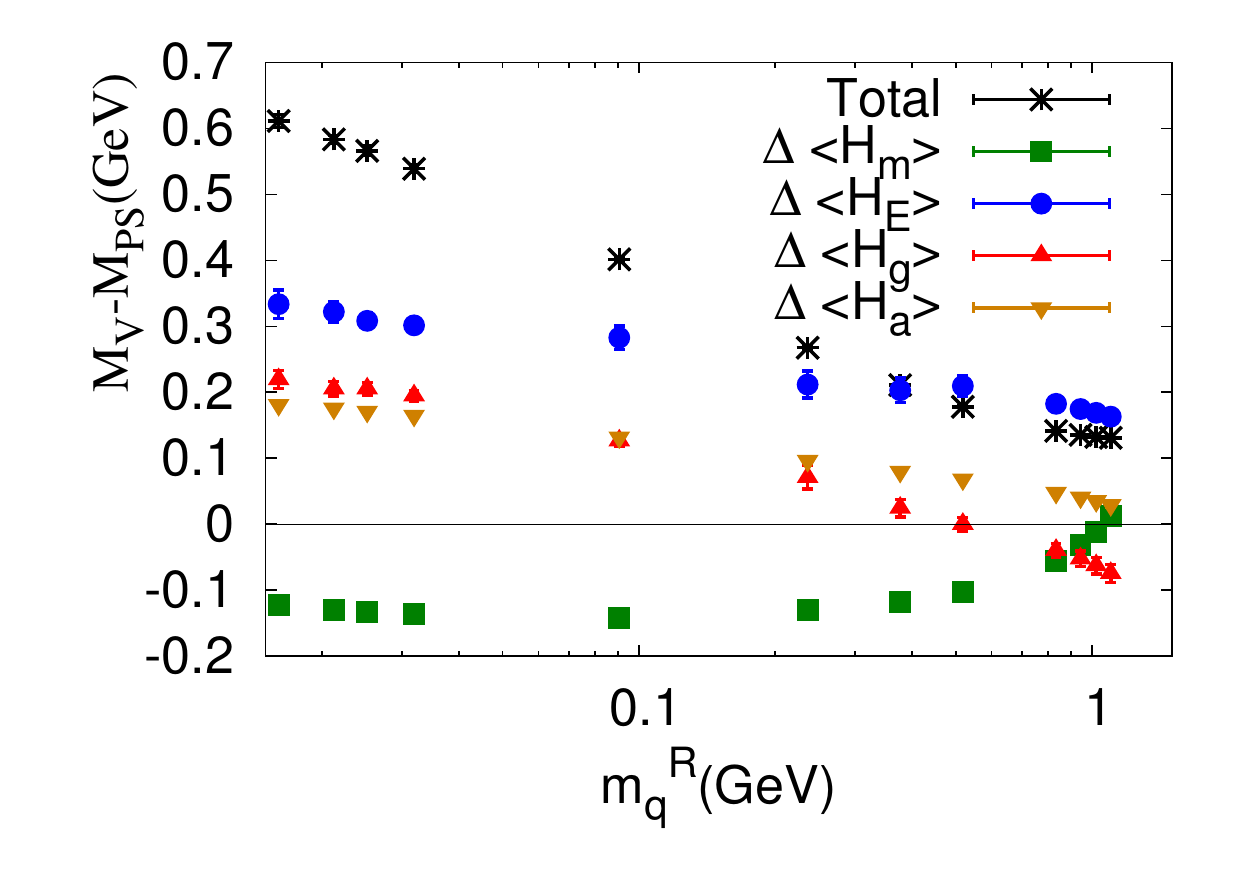}
  \includegraphics[scale=0.6]{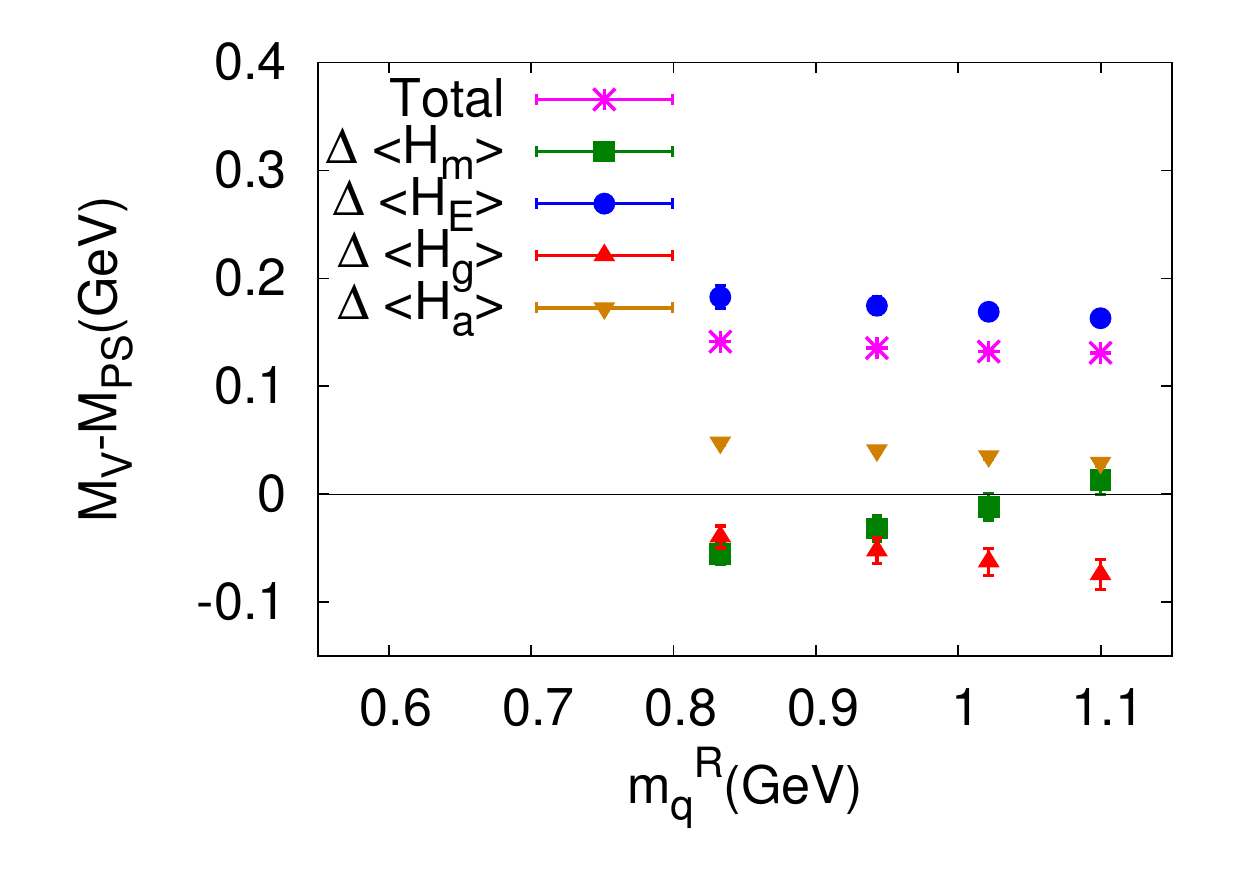}
\caption{Contributions to hyperfine splitting for the entire quark mass region (the top panel) and an enlarged plot in the charm quark region (the bottom panel). }\label{fig:hfs1}
\end{figure}

\begin{figure}[]
  \includegraphics[scale=0.6]{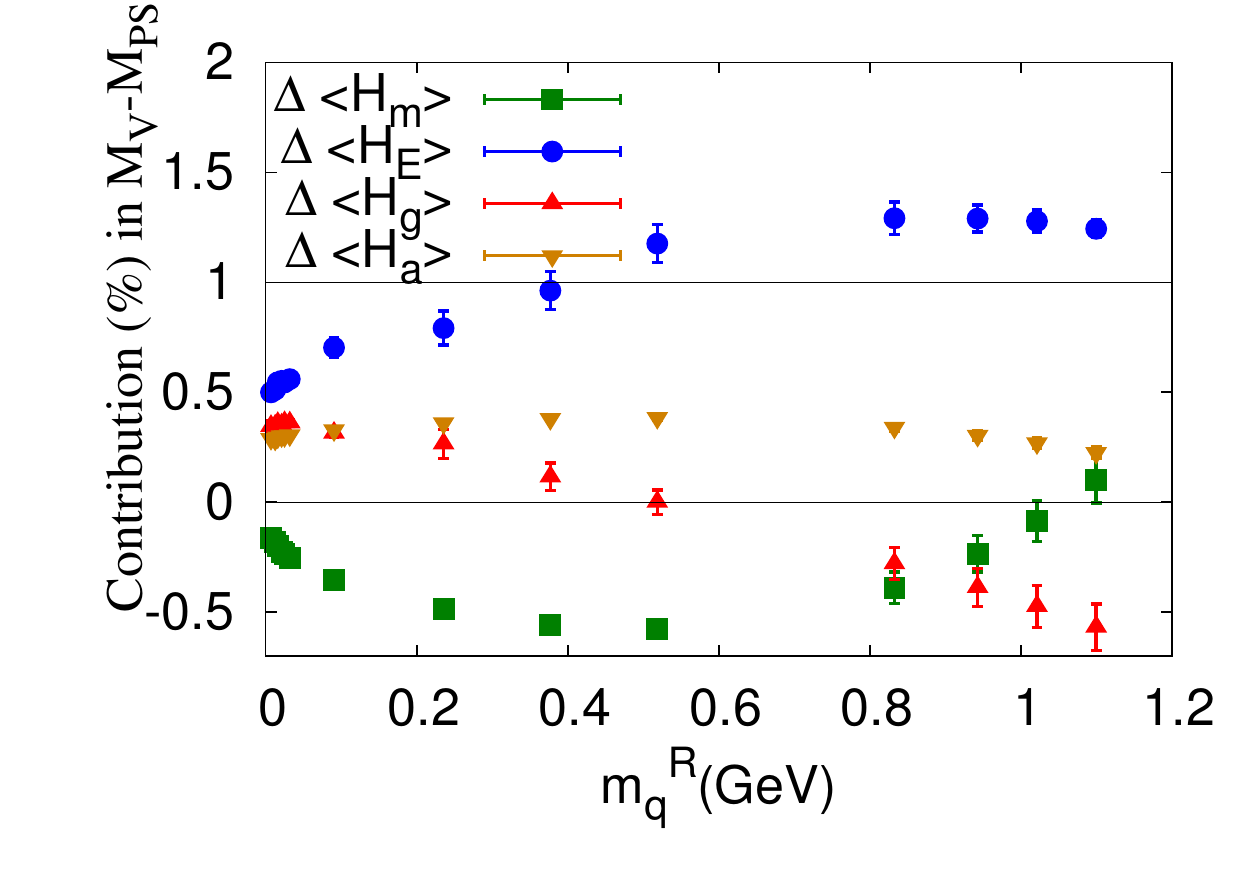}
\caption{Contributions to the hyperfine splitting, as portions of the total. In the charm quark mass region, $\Delta  \langle H_a\rangle$ (orange reverse triangle) contributes 1/4 since the one from the mass term (green square) is nearly zero. At the same time the contribution from the glue energy difference (red triangle) turns negative, largely canceling the positive $\Delta \langle H_a\rangle$, and leaves the major part of the hyperfine splitting as due to the quark energy difference $\Delta \langle H_E\rangle$ (blue dot).}\label{fig:hfs2}
\end{figure}

As seen in Fig.~\ref{fig:glue_ps1} and Fig.~\ref{fig:glue_v1}, when the valence quark mass increases, the quark energy contribution also increases while both the glue energy and trace anomaly decrease. To study the consequence
of this behavior, we examine the hyperfine splitting of charmonium, we plot in Fig.~\ref{fig:hfs1} the difference of the quark and glue components between the V and PS mesons as a function of the quark mass. For charmonium
with $m_q^R$ at 1 GeV, $\Delta \langle H_m\rangle$ is consistent with zero. Therefore, as in Fig.~\ref{fig:hfs2}, $\Delta  \langle H_a\rangle$ gives 1/4 of the hyperfine splitting from the trace anomaly equation 
Eq.~(\ref{eq:trace}). On the other hand, $\Delta \langle H_g\rangle$ turns negative in the charm mass region and
largely cancels out the positive $\Delta  \langle H_a\rangle$. As a result, the major part of the hyperfine splitting is
due to the quark energy difference $\Delta \langle H_E\rangle$. This seems to be consistent with the
potential model picture where the charmonium hyperfine splitting is attributable to the spin-spin interaction of
the one glue-exchange potential. Higher precision calculation is needed to confirm this.

\section{Summary}\label{sec:summary}

In summary, we have directly calculated the quark components of the pseudoscalar and vector meson masses with 
lattice QCD. The glue field energy and the anomaly components are extracted from the mass relations from the Hamilton and the trace anomaly. 
 We have estimated the systematic errors due to the neglect of the disconnected insertions and the use of equation of motion. 
From our exploratory study, we confirmed that there are significant contributions from the glue components in light mesons. Throughout the valence quark mass range below the charm, the quark mass dependence of the V meson mass comes almost entirely from $\langle H_m\rangle$, which is linear in the valence quark mass; whereas, 
$\langle H_E\rangle $, $\langle H_a\rangle$ and $\langle H_g\rangle$ are almost constant. 
We also find that the hyperfine splitting between $J/\Psi$ and $\eta_c$ is dominated by the
quark energy term. 

\begin{figure}[]
  \includegraphics[scale=0.6]{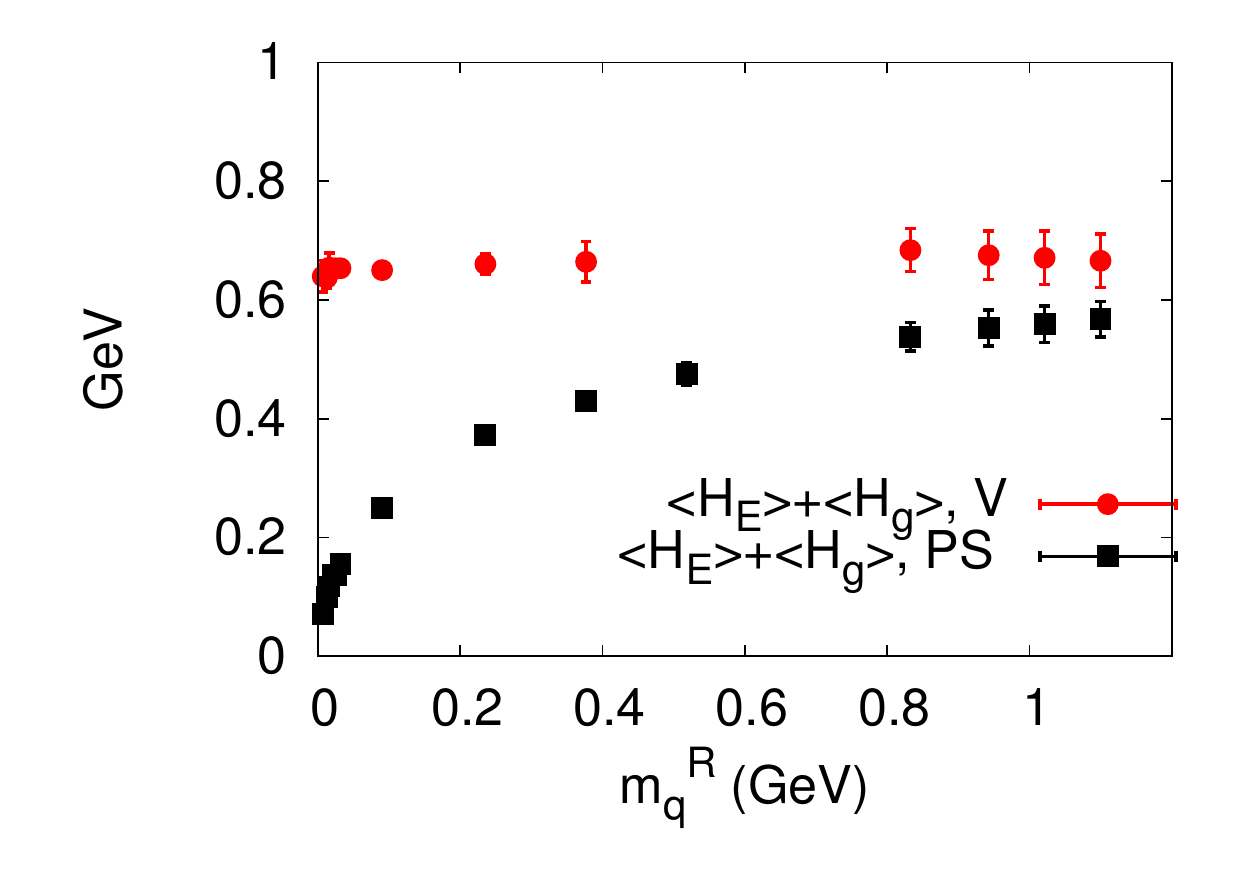}
\caption{The combined quark/glue energy contribution to the PS/V meson mass.}\label{fig:energy}
\end{figure}

In the decomposition equation Eq.~(\ref{eq:T44_new}), 
each of the first two terms and the sum of the last two terms are separately scale and renormalization scheme independent, while each of the last two terms is not. 
For reference, the scale independent combination of the quark 
and glue energy contributions to the PS/V meson mass are plotted in Fig.~\ref{fig:energy}. We can see that the combined energy in the V meson is almost a constant ($\sim$ 650 MeV), and the one in the PS meson becomes close to that value only in the charm quark region.

For future studies, we will perform calculations with the physical sea quark masses and will calculate the glue field energy and trace anomaly contributions directly.

\section*{Acknowledgments}

We thank RBC and UKQCD Collaborations for providing us their DWF gauge configurations. This work is supported in part by the National Science Foundation of China (NSFC) under Grants
No. 11075167, No. 11105153, and No. 11335001, and also by the U.S. DOE Grant No. DE-FG05-84ER40154.
Y.C. and Z.L. also acknowledge the support of NSFC and DFG through funds provided to the
Sino-German CRC 110 ``Symmetries and the Emergence of Structure in QCD". M. G. and Z. L. are partially supported by the Youth Innovation Promotion Association of CAS (2015013, 2011013).

\end{document}